\documentclass[12pt,a4paper]{article}
\usepackage[utf8]{inputenc}
\usepackage[english]{babel}
\usepackage{amsmath}
\usepackage{amstext}
\usepackage{amscd}
\usepackage{amsfonts}
\usepackage{amssymb}
\usepackage{mathrsfs}

\newtheorem{theorem}{Theorem}

\newtheorem{proof*}{Proof}

\usepackage[pagebackref,draft=false]{hyperref}
\hypersetup{colorlinks,
linkcolor=myrefcolor,
citecolor=mycitecolor,
urlcolor=myurlcolor}

\usepackage[capitalize]{cleveref}
\usepackage{cite}
\usepackage{caption}
\usepackage{etaremune}

\usepackage{xcolor}
\definecolor{myurlcolor}{rgb}{0,0,0.4}
\definecolor{mycitecolor}{rgb}{0,0.5,0}
\definecolor{myrefcolor}{rgb}{0.5,0,0}
\usepackage{graphicx}
\usepackage{tikz}
\usepackage{tikz-cd}
\usetikzlibrary{cd}
\usetikzlibrary{decorations.pathmorphing,shapes}

\usepackage{etoolbox}
\usepackage{enumitem} 
\usepackage[]{latexsym}
\usepackage{braket}
\usepackage{caption}
\usepackage[utf8]{inputenx}
\usepackage[T1]{fontenc}
\usepackage{lmodern}
\usepackage{textcomp}
\usepackage{microtype}
\usepackage{totcount}
\usepackage{blindtext}

\newcommand{\be}{\begin{equation}}
\newcommand{\ee}{\end{equation}}

\title{Quantum Tomography and Schwinger's Picture of Quantum Mechanics}
\date{}

\author{F. M. Ciaglia $^{1,6}$ \href{https://orcid.org/0000-0002-8987-1181}{\includegraphics[scale=0.7]{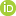}}, F. Di Cosmo$^{1,2,7}$ \href{https://orcid.org/0000-0003-0256-5913}{\includegraphics[scale=0.7]{ORCID.png}}, A. Ibort$^{1,2,8}$ \href{https://orcid.org/0000-0002-0580-5858}{\includegraphics[scale=0.7]{ORCID.png}},\\ G. Marmo$^{3,4,9}$ \href{https://orcid.org/0000-0003-2662-2193}{\includegraphics[scale=0.7]{ORCID.png}} \\
\footnotesize{$^{1}$\textit{Departamento de Matem\'aticas, Univ. Carlos III de Madrid, Legan\'es, Madrid, Spain}} \\
\footnotesize{$^{2}$\textit{ ICMAT, Instituto de Ciencias Matem\'{a}ticas (CSIC-UAM-UC3M-UCM)}}\\
\footnotesize{$^{3}$\textit{ INFN-Sezione di Napoli, Naples, Italy}} \\
\footnotesize{$^{4}$\textit{ Dipartimento di Fisica ``E. Pancini'', Universit\`a di Napoli Federico II,  Naples, Italy}} \\
\footnotesize{$^{6}$\textit{ e-mail: \texttt{fciaglia[at]math.uc3m.es}}} \\
\footnotesize{$^{7}$\textit{ e-mail: \texttt{fcosmo[at]math.uc3m.es}}} \\
\footnotesize{$^{8}$\textit{ e-mail: \texttt{albertoi[at]math.uc3m.es}}} \\
\footnotesize{$^{9}$\textit{ e-mail: \texttt{gmarmo[at]unina.it}}} 
}

\begin{document}
\maketitle

\begin{abstract}
In this paper the problem of tomographic reconstruction of states is investigated within the so-called Schwinger's picture of Quantum Mechanics in which a groupoid is associated with every quantum system. The attention is focused on spin tomography: In this context the groupoid of interest is the groupoid of pairs over a finite set. In a nutshell, this groupoid is made up of transitions between all possible pairs of outcomes belonging to a finite set. In addition, these transitions possess a partial composition rule, generalizing the notion of groups. The main goal of the paper consists in providing a reconstruction formula for states on the groupoid-algebra associated with the observables of the system. Using the group of bisections of this groupoid, which are special subsets in one-to-one correspondence with the outcomes, a frame is defined and it is used to prove the validity of the tomographic reconstruction. The special case of the set of outcomes being the set of integers modulo n, with n odd prime, is considered in detail. In this case the subgroup of discrete affine linear transformations, whose graphs are linear subspaces of the groupoid, provides a \textit{quorum} in close analogy with the continuos case.  
\end{abstract}

\section{Introduction}
The relation between classical and quantum mechanics is a long-standing question in Quantum Theory. Since the seminal works by Bohr, Dirac and Weyl, classical mechanics was meant to be classical Hamiltonian mechanics, due to the direct analogy between Poisson brackets and commutators and between Hamilton and Heisenberg equations. Many characteristic features of Quantum Theories have been understood using methods from symplectic and Poisson geometry and one fruit of this investigation was the geometric quantization program by Kirillov, Konstant and Souriau \cite{Konstant-1970,Kirillov-2001,Souriau-1966}. However, quantization is only one side of the correspondence between classical and quantum mechanics and, if we insist on viewing quantum theories as being ``more fundamental'', quantization is not the most appropriate tool to obtain a deeper understanding of quantum phenomena. Indeed a fundamental observation is the fact that classical behaviours of physical systems are obtained from quantum ones by taking a suitable limit. Consequently, there must always be a \textit{dequantization} procedure describing such a limit and classicality should appear in this way. This was the rationale behind the generalization of Weyl-Wigner correspondence\cite{Weyl-1927,Wigner-1997} to the Quantizer-Dequantizer formalism proposed by some of the authors \cite{MMM-2000,CDIM-2017}. Koopman's paper \cite{Koopman} can be considered a seminal work in this direction, since for the first time classical mechanics was described using the same formalism of Schr\"{o}dinger's formulation of Quantum Mechanics. Once a unified description for quantum and classical systems exists the problem of the emergence of classicality can be approached more directly. Despite such an unified approach, quantum and classical systems have very different properties and being able to distinguish them is a crucial problem which is to be faced especially for the development of Quantum Technologies \cite{CWZ-2010,HTFXLG-2020,WSR-2019}. In particular, being able to discriminate between quantum and classical states is one of the fundamental issues in current quantum computing \cite{PZ-2017,CCDG-2005,FR-2016}. 

The modern notion of state has its roots back to von Neumann's ``Gesamtheiten'' \cite{VonNeumann-1927}. As stated by Leonhardt, \textit{``knowing the state means knowing the maximally available statistical information about all physical quantities of a physical object''} \cite{Leonhardt1997}. This corresponds to the knowledge of the probability distributions associated with all the possible observables of a physical system \cite{CWZ-2010}. While classically a state is a single probability distribution on the phase space of the system, in the quantum case there are infinitely many different probability distributions associated with every quantum state and it is not possible to obtain information about all of them simultaneously. However, a big amount of this information is redundant, and, in many instances, a quantum state can be actually reconstructed from the knowledge of a finite number of probability distributions (certainly for a system with a finite number of degrees of freedom). This reconstruction problem is the fundamental question of Quantum Tomography. The problem of Quantum Tomography originates from a question raised by Pauli \cite{Pauli-1933} and answered in the negative by Bargmann \cite{Reichenbach-1998}: Is it possible to reconstruct any wave function from the probability distributions associated with position and momentum operators? Since then, Quantum Tomography has rapidly evolved and it has become a fundamental chapter in Quantum Information Theory: in particular, here one can distinguish exact tomography from real tomography (see for instance the books \cite{NC-2000,PR-2004}). In exact tomography no source of error is considered and the exact reconstruction of the original state can be achieved in a finite amount of time. On the other hand, real tomography deals with physical situation where the occurrence of errors needs to be taken into account in order to get meaningful results (see for instance the papers \cite{REK-2004,JKMW-2001} where real tomographic procedures for single and multiple qubits are described). In this case, an effective estimation of the state of the system can be obtained only after a post-measurement process, and adaptive procedures in this phase are under investigation (see for instance the recent algorithm \cite{GRSTBMTM-2021}). Additionally, the features of the involved state may suggest more efficient procedures which can be adopted, like tomography via compressed sensing illustrated in \cite{GLFBE-2010,FGLE-2012} which is suited for the reconstruction of low rank states. In this paper we will focus only on exact tomography, whereas the problem of state estimation and real tomography will be dealt in forthcoming works.     

There exist many different procedures for the exact tomographic reconstruction of a state. From one point of view, it is possible to perform different measures on copies of the state obtained from the same ensemble: each measurement will provide the experimenter with a probability distribution and the knowledge of a sufficient number of probabilities allows for the reconstruction of the state. From this point of view, if we have a quantum system with a finite number of degrees of freedom, say n, the maximum amount of information about the ensemble can be got if one performs n+1 measurements which are mutually unbiased \cite{WF-1989,Ivonovic-1981}, i.e., the associated orthonormal bases $\left\lbrace \mid e^{(r)}_j \rangle \right\rbrace$, with $r=1,2, \cdots,n+1$ and $j=1,2,\cdots,n$ satisfy the property $|\langle e^{(r)}_j|e^{(s)}_k\rangle|^2 = 1/n $ whenever $r\neq s$. 

On the other hand, instead of considering a set of unbiased measurements, one could consider a more efficient tomographic reconstruction via Symmetric Informationally Complete-POVMs (SIC-POVMs), i.e., a positive operator valued measure whose statistics allows to reconstruct the state\cite{Busch-1991,RBKSC-2004}. A SIC-POVM for a n-dimensional quantum system consists of $n^2$ positive operators, each one being a multiple of a projector $P_j = |\psi_j \rangle \langle \psi_j|$, such that their sum is the identity operator and the adjective symmetric refers to the property that $|\langle \psi_j|\psi_k\rangle|^2 = 1/(n+1)$, for any $j,k$. This number, however, can be reduced in the particular instance of pure states tomography, and the minimal number of vectors in an Informationally Complete POVM is 2N \cite{FSC-2005}. 

The existence of mutually unbiased orthonormal bases (MUBs) and SIC-POVMs in every dimension is still an open question and only partial results are known. For instance MUBs have been proven to exist in every dimension power of a prime number \cite{WF-1989,Ivonovic-1981}, whereas, using group theory (S)IC-POVMs have been found in many dimensions (see for instance \cite{RBKSC-2004,DPS-2004} for some first examples of these covariant POVMs) but an existence theorem for every $n$ is still lacking.  

In this paper, we exploit the so-called Schwingers picture of Quantum Mechanics  \cite{CIM-2018,CDIM-2020a} based on groupoids, in order to get reconstruction formulae for a state of a quantum system with a finite number of degrees of freedom. In this context, to every quantum system there is associated a groupoid $\mathcal{G}(\Omega)$ over a finite set $\Omega$ and a von Neumann algebra $\mathcal{V}(\mathcal{G})$ which is generated by suitable functions on $\mathcal{G}(\Omega)$ endowed with a convolution product. Moreover, $\mathcal{V}(\mathcal{G})$ is represented on the Hilbert space $\mathcal{H}(\mathcal{G})$ of square-integrable functions on $\mathcal{G}(\Omega)$. Using the notion of bisection \cite{Mackenzie-2005,CDIMS-2021} we show that it is possible to build a frame for $\mathcal{H}(\mathcal{G})$ when $\mathcal{G}(\Omega)$ is the pair groupoid of a set $\Omega$ with finite cardinality, say $n$. Then, using the decomposition of the identity associated with this frame we are able to provide a reconstruction formula for any state.

A fundamental ingredient of this approach is that characteristic functions supported on bisections determine a unitary representation of the permutation group $\mathbb{S}_n$, for generic values of $n$. However, contrarily to the situation investigated in many previous works \cite{DMP-2003,DMP-2000,CDDL-2000}, we do not use an irreducible representation, but the fundamental one. Moreover, in the case of odd prime dimension, a subgroup of the whole symmetric group $\mathbb{S}_n$ is sufficient to produce a frame, i.e., the group of discrete affine linear transformations (see Sect.\ref{sec 4.3}). It is interesting to notice that the associated graph is the discrete analogue of a line and one can interpret the reconstruction formula as the discrete analogue of the generalized Radon transform using the set of all Lagrangian subspace of a symplectic vector space \cite{MMN-2014}. Despite of this nice geometric analogy, the frame which is constructed in this paper is not tight and consequently it cannot be directly used to construct a (S)IC-POVM. Nevertheless, generalizations to address this problem can be conceived.

Another relevant aspect consists in the fact that using a different approach directly inspired by classical Radon transform, we derive a tomographic approach which is similar to the one based on the so called (generalized) Pauli group in the discrete phase space \cite{GHW-2004,KMR-2006,ABC-2008}, which has been shown to have a deep connection with the notion of MUBs and IC-POVMs \cite{KRBS-2007,SRK-2010}. Indeed, in these works displacement operators are constructed satisfying the relations $XY = \omega YX$ ($\omega = \exp(i2\pi/n)$) and using their powers a sufficient number of mutually unbiased orthonormal bases is obtained. These operators can be grouped into representations of the group of integers modulo $n$, each one providing a basis of the Hilbert space $\mathbb{C}^n$. In the rest of this paper, instead of using a representation of the Pauli group and the associated displacement operators, we will prove that the discrete analogue of a Lagrangian submanifold can be obtained for the pair groupoid $\mathcal{G}(\Omega)=\Omega\times\Omega$, with $\Omega$ the finite field of order $n$ and $n$ odd prime. In this case, a bisection defines an immersion map of $\Omega$ within $\mathcal{G}(\Omega)$ and the bisections associated with the graphs of linear maps form the analogue of linear subspaces over which we integrate the function associated with a state. In this sense the procedure outlined in this paper is closely related to the classical idea of tomographic reconstruction, even if the setting is purely quantum. In particular, we will also find a family of $n+1$ orthonormal bases, but they are not mutually unbiased. A way to recover the previously mentioned mutually unbiased bases will be presented in Sec.\ref{sec 4.3}.

Apart from introductory and conclusive sections, the paper contains three sections. In Sec.\ref{sec.2} the ingredients of a mathematical description of a physical system are provided, to set the stage for the unified approach to classical and quantum mechanics previously mentioned. Using such frame, we briefly recall the main ingredients of the different formulations of Quantum Mechanics. In Sec.\ref{Q-D_section} we present the Weyl-Wigner correspondence and its connection to Quantum Tomography. Finally, in Sec.\ref{sec.4 spin tomography} we illustrate a tomographic reconstruction procedure adapted to the groupoid picture of Quantum Mechanics. The two cases of generic $n$ and $n$ odd prime, are presented and the peculiarities of the second case are highlighted.

\section{Mathematical description of physical systems}\label{sec.2}
Every mathematical description of a physical system requires the introduction of some basic ingredients:
\begin{itemize}
\item a set $\mathcal{O}$ whose elements are the observables of  the theory;
\item a set $\mathfrak{S}$ whose elements are the states of the theory (usually, it is a convex set);
\item a statistical interpretation provided by a map, $\mu \, \colon \mathfrak{S}\times \mathcal{O}\,\rightarrow \, \mathscr{M}\left( \mathbb{R} \right)$, where $\mathscr{M}\left(\mathbb{R}\right)$ denotes the space of probability measures on the real line;
\item a dynamical evolution law provided by a family of consistent evolution operators, 
\begin{equation}
U_{t,s}\,\colon\, \mathfrak{S}\,\rightarrow \,\mathfrak{S}\,,\quad t,s\in \mathbb{R}_+\,,\quad U_{t_2,t_1} = U_{t_2,s}\circ U_{s,t_1}\,;
\end{equation}
\item an algorithmic rule to compose physical systems.
\end{itemize}  
Concerning the evolution map, with additional requirements about its diferentiability, it can be determined by a differential equation. The probability map $\mu$, instead, is interpreted as follows: the number $\mu(\rho, A)(E)$, with $\rho \in \mathfrak{S}$, $A\in \mathcal{O}$, and $E$ a measurable subset of $\mathbb{R}$, is the probability that a measurement of the observable $A$ gives a result in the subset $E$. 

According to this scheme, we can illustrate the different formalisms of both Quantum and Classical Mechanics. Indeed, in Classical Mechanics a physical system is described by a pair $(\mathcal{M}, \omega)$, where $\mathcal{M}$ is a suitable manifold and $\omega$ is a symplectic 2-form. The set of observables is given by a set of measurable real-valued functions $\mathscr{F}(\mathcal{M}) =: \mathcal{O}$, which also forms an associative commutative algebra. The set of states is given by the family of probability measures $\mathfrak{S}:= \mathscr{P}(\mathcal{M})$ on the manifold $\mathcal{M}$. Given a state $\nu\in \mathfrak{S}$ and an observable $f\in \mathcal{O}$, the probability map $\mu$ is provided by the measure, $\mu(\nu , f)(E) = \nu(f^{-1}(E))$, according to which the observables are interpreted as random variables on a probability space. The evolution map is a one-parameter group of symplectic transformations, $\varphi_{t}\,\colon\,\mathcal{M}\,\rightarrow\,\mathcal{M}$ possessing additional regularity properties. In the case of differentiable one-parameter group of symplectic transformations we recover an infinitesimal description given by the well-known Hamilton equations. Finally, if we have two physical system $(\mathcal{M}_1,\omega_1 )$ and $(\mathcal{M}_2, \omega_2)$, respectively, we define their composition as the system $(\mathcal{M}_1\times \mathcal{M}_2, \omega_1 \oplus \omega_2)$.
 
On the other hand, we have different formulations, also called ``pictures'', for Quantum Mechanics. In the standard picture, the departure point is the association of a Hilbert space $\mathcal{H}$ with a quantum system. Then, the observables are represented as self-adjoint operators on $\mathcal{H}$; the states of the system are identified with non-negative linear operators on $\mathcal{H}$ with unit trace (density operators), and the extreme points (pure states) of the space of states $\mathfrak{S}$ are rank-one projectors which are in 1-1 correspondence with rays in $\mathcal{H}$ (elements of the complex projective space $\mathbb{P}(\mathcal{H})$). The evolution map $U_t\,\colon\, \mathcal{H}\,\rightarrow \,\mathcal{H}$ is a one-parameter group of unitary transformations of the Hilbert space $\mathcal{H}$. At the infinitesimal level the evolution is determined by the Schr\"{o}dinger equation. As for the probabilistic interpretation, given a self-adjoint operator $A\in \mathscr{B}(\mathcal{H})$ we have its spectral decomposition $A = \int_{\sigma(A)} \lambda \, \mathrm{d}\Pi_{\lambda}$, where $\Pi_{\lambda}$ is a projection-valued measure on the spectrum $\sigma \subset \mathbb{R}$ of the operator $A$. Then, given a pure state $\rho_{\psi}$ and an observable $A$ we have the probability map 
\begin{equation}
\mu (\rho_{\psi}, A) (E) = \int_{E} \frac{ \, \left\langle \psi \,|\, \mathrm{d} \Pi_{\lambda} \psi \right\rangle}{\left\langle \psi \,|\, \psi \right\rangle}\,.
\end{equation}
The quantity $e_{\psi}(A) := \frac{\left\langle \psi \, | \, A\psi \right\rangle}{\left\langle \psi \, | \, \psi \right\rangle}$ is interpreted as the expectation value of the observable $A$ on the state associated to the vector $|\left. \psi \right\rangle \in \mathcal{H}$, whereas $e_{\psi}(A^2) - (e_{\psi}(A))^2$ is the corresponding dispersion. The probabilistic interpretation is straightforwardly extended to mixed normal states using the trace on $\mathscr{B}(\mathcal{H})$. Regarding the composition, quantum systems are composed according to von Neumanns composition rule, that is, the Hilbert space of the total system is the tensor product of the composing systems.

It was noticed almost immediately that the Hilbert space formalism was not suited only to Quantum Mechanics but it could be adapted to describe also Classical Mechanics. This fundamental observation was elaborated in 1931 by Koopman in the paper ``Hamiltonian systems and transformations in Hilbert space'' \cite{Koopman} and it was the departing point of an extremely active field of research on the relation between Classical and Quantum Mechanics. Then, according to the previous scheme, given a classical Hamiltonian system $(\mathcal{M}, \, \omega)$, with $\mathrm{dim}(\mathcal{M}) = 2n$, we can associate the Hilbert space $\mathcal{H}= \mathcal{L}^2(\mathcal{M}, \nu)$ where $\nu$ is the measure determined by the volume form $\mathrm{vol}_{\mathcal{M}}=\omega^n$. The set of observables, $\mathcal{O}$, is formed by multiplication operators on $\mathcal{H}$, and the states are probability measures absolutely continuous with respect to the measure $\nu$. The evolution map is given by the one-parameter group of unitary transformations, $(U_t\psi)(m)=\psi(\varphi_t(m))$, where $m\in\mathcal{M}$ and $\varphi_t$ is a one-parameter group of symplectomorphisms of $\mathcal{M}$. The Hilbert space associated with a composite system is the tensor product of the subsystems. Therefore, the same formalism originated in Quantum Mechanics can be used also for the description of classical systems. The need for a formalism capable of accommodating the description of both classical and quantum systems is now considered as an important feature of a theoretical description of physical systems. 

Another way of looking at this feature comes from the Heisenberg picture of Quantum Mechanics, a description which is dual to the Schr\"{o}dinger's one, in the sense that the focus is on the space of observables, instead of the set of states. In this case every physical system is associated with a $C^*$-algebra \cite{Blackadar-2006}, say $\mathcal{A}$, whose real elements, i.e. invariant under the involution $\ast \,\colon \, \mathcal{A}\,\rightarrow \, \mathcal{A}$ of the algebra, are interpreted as the observables of the system. The set $\mathcal{O}$ of observables is not a subalgebra, since the product of two real elements does not need to be real. However, it can be endowed with a Jordan algebra structure, whose structure is connected with fundamental questions of Quantum Information Theory which are currently under investigation (see \cite{CJS-2020,FFIM-2012}). The set $\mathfrak{S}$ of states is the set of positive linear functionals of norm one, where the norm is the norm of the dual Banach space $\mathcal{A}^*$. The statistical interpretation is recovered by noticing that the functional calculus for $C^*$-algebras \cite{Blackadar-2006} extends the functional calculus on the algebra $\mathscr{B}(\mathcal{H})$ and the probability map is given by
\begin{equation}
\mu(\rho, A) (E) = \rho(\int_E\,\mathrm{d}\Pi_{\lambda})\,,
\end{equation}  
where $\Pi_{\lambda}$ is the projection-valued measure on the spectrum of a real element of a $C^*$-algebra. Therefore, we get the expectation value map $e_{\rho}(A) = \rho(A)$. The evolution map, instead, is given by a one-parameter group $\varphi_t$ of automorphisms of the algebra $\mathcal{A}$ which at the infinitesimal level is generated by a derivation of the algebra. In the particular instance that $\varphi_t$ consists of inner automorphisms of the algebra we recover at the infinitesimal level the Heisenberg equation. Finally, given two physical systems described by the algebras $\mathcal{A}_1$ and $\mathcal{A}_2$, respectively, the composite system is associated with the tensor product algebra, where a suitable completion has to be chosen for the tensor product in order to get a $C^*$-algebra.   
In this context, classical systems arise when the $C^*$-algebra is Abelian while the standard Hilbert space formalism of Quantum Mechanics is essentially recovered when the $C^*$-algebra is $\mathcal{B}(\mathcal{H})$.

\section{Weyl-Wigner correspondence}
\label{Q-D_section}
In order to contextualize our approach to tomography let us recall briefly the Weyl-Wigner correspondence and the description of Quantum Mechanics on phase space. Let us start from considering an Abelian vector group $\mathcal{M}$ with a symplectic structure $\omega$ invariant with respect to the action of the group on itself. A Weyl system is a map $W\,\colon\,(\mathcal{M},\,\omega)\,\rightarrow \, \mathcal{U}(\mathcal{H})$ satisfying 
\begin{equation}\label{eq:Weyl-commutation relations}
W(v_1)W(v_2)W^{\dagger}(v_1)W^{\dagger}(v_2)=\mathrm{e}^{i\omega(v_1,\,v_2)}\mathbf{1}\,,\quad v_1,\,v_2\in \mathcal{M}\,,
\end{equation}
where $\mathcal{U}(\mathcal{H})$ is the group of unitary transformations of a Hilbert space $\mathcal{H}$. In other words, a Weyl map defines a projective representation on the Hilbert space $\mathcal{H}$ of the Abelian group $\mathcal{M}$.

A standard way to construct such a representation is to consider a Lagrangian subspace $L\subset \mathcal{M}$, so that $\mathcal{M} \cong \mathbf{T}^*L$. Then we introduce the Hilbert space $\mathcal{H}_L=\mathcal{L}^2(L,\mu)$, where $\mu$ is a measure invariant under the action of the group itself. Therefore, $\mathcal{M}$ can be described by the coordinate functions $(x,f)$, where $x$ belongs to $L$ and $f$ denotes a linear functional in the dual Abelian group $L^*$. Then, we can define a Weyl map as follows
\begin{equation}
\begin{split}
U\,\colon\,L\,\rightarrow\,\mathcal{U}(\mathcal{H}_L) \quad (U(x)\psi)(y)= (\mathrm{e}^{ix\hat{P}}\psi)(y)=\psi(y+x)\\
V\,\colon\,L^*\,\rightarrow\,\mathcal{U}(\mathcal{H}_L) \quad (V(f)\psi)(y)= (\mathrm{e}^{if\hat{Q}}\psi)(y)=\mathrm{e}^{if(y)}\psi(y)
\end{split}
\end{equation}
so that the commutation relations in Eq.\eqref{eq:Weyl-commutation relations} hold. Eventually, by using the symplectic Fourier transform $\tilde{A}$ of any function $A$ on $\mathcal{M}$
\begin{equation}
A(q,p)=\int_{\mathcal{M}}\tilde{A}(x,f) \mathrm{e}^{i(p(x)-f(q))}\,\mathrm{d}x\,\mathrm{d}f\,,
\end{equation}
we can build an operator on the Hilbert space $\mathcal{H}_L$ as follows
\begin{equation}
A(\hat{Q},\hat{P}) = \int_{\mathcal{M}} \tilde{A}(x,f)\mathrm{e}^{i(x\hat{P}-f\hat{Q})}\,\mathrm{d}x\,\mathrm{d}f\,.
\end{equation}
It is also possible to invert the previous relation, so that given an operator on $\mathcal{H}_L$ we can associate a function on the group $\mathcal{M}$. In particular, if $F(x,x')$ is the position representation of an operator on $\mathcal{H}_L$ we have the following transformation
\begin{equation}
\mathcal{W}_F(q,p)=\int_{L} F\left(q+\frac{x}{2},\,q-\frac{x}{2}\right) \mathrm{e}^{-ixp}\,\mathrm{d}x\,,
\end{equation}
which is called Wigner transform. 

It is possible to look at the previous construction from a different perspective. To this aim let us first make a short digression. For the sake of simplicity let us consider a finite set $\mathcal{S}$ of elements, say $s_1,\,s_2,\,\cdots\,,\,s_n$ and take the formal linear combinations
\begin{equation}
A_f = \sum_{j=1}^{n}f(s_j)s_j \,,
\end{equation}    
where $f$ is a complex-valued function on $\mathcal{S}$ (let us note here that the function $f$ does not need to be complex-valued but may be real, quaternionic or even p-adic). The set of all these formal combinations forms a vector space, say $\mathcal{V}(\mathcal{S})$. Then, if the set $\mathcal{S}$ is endowed with a binary composition rule
\begin{equation}
\beta\,\colon\,\mathcal{S}\times\mathcal{S}\,\rightarrow\,\mathcal{S}\,,
\end{equation}
the vector space $\mathcal{V}(\mathcal{S})$ becomes an algebra with respect to the product 
\begin{equation}
A_f \cdot A_g = \left( \sum_{j=1}^{n}f(s_j)s_j \right) \cdot \left( \sum_{k=1}^{n}g(s_k)s_k \right) = \sum_{j=1}^{n}\sum_{k=1}^{n}f(s_j)g(s_k)\beta(s_j,s_k)\,.
\end{equation}
If the internal composition $\beta$ determines a group structure on the set $\mathcal{S}$ the algebra $\mathcal{V}(\mathcal{S})$ is an associative algebra called the group-algebra of the group $\mathcal{S}$. The passage to the continuous case can be performed by introducing a measure invariant with respect to the action of the group on itself and replacing the sum with an integral over the group \cite{Folland-2016}. Then, if the group has a representation $U\,\colon\,\mathcal{S} \,\rightarrow\,\mathcal{B}(\mathcal{H})$ on some Hilbert space $\mathcal{H}$, the elements of the group-algebra are represented as operators on $\mathcal{H}$ according to $\hat{A}_f = \sum_{j=1}^{n} f(s_j)U(s_j)$.

Coming back to the Abelian group $\mathcal{M}$ and the maps $U$ and $V$, its central extension by means of a 2-cocycle $\omega$ defines the \textit{Heisenberg-Weyl} group and the Weyl map is a representation of this group on the Hilbert space $\mathcal{H}_L$ \cite{Kirillov2004}. Therefore, the previous construction can be now reinterpreted by saying that the Weyl map defines a representation of the group-algebra of the Heisenberg-Weyl group on the Hilbert space $\mathcal{H}_L$. 

Summarizing the previous discussion we have seen that according to the Weyl-Wigner correspondence it is possible to describe Quantum Mechanics using functions on an auxiliary space $\mathcal{M}$ and then operators on a suitable Hilbert space can be reconstructed from functions on $\mathcal{M}$. This procedure provides an opposite point of view with respect to Koopman's one: we can analyse quantum systems also using tools coming from Classical Mechanics. However, due to the difference between Quantum and Classical Mechanics, the properties of the classical-like objects deriving from the quantum ones are not as one would expect in the description of a classical system. For instance, the function that one associates to the operator representing a normal quantum state via the Wigner map does not need to be positive. This undesired consequence can be avoided introducing the notion of tomogram (for an introductory description see, for instance \cite{IMMS-2009}). 

In a sentence, the classical problem of tomography consists in the reconstruction of a function defined over a certain space knowing its value on specific subsets. For instance, in the case of tomography on $\mathbb{R}^2$ we have a function $f\,\colon\,\mathbb{R}^2\,\rightarrow\,\mathbb{R}$ and one wants to reconstruct it just knowing its mean value on straight lines in $\mathbb{R}^2$. More specifically, if we have a nonnegative normalized and smooth function $f\,\colon\,\mathbb{R}^2\,\rightarrow\,\mathbb{R}$, i.e., $f(q,p)\geq 0$ $\forall(q,p)\in \mathbb{R}^2$ and $\int_{\mathbb{R}^2}f(q,p)\,\mathrm{d}q \,\mathrm{d}p = 1$, we define its tomogram (also called its Radon transform) as 
\begin{equation}
\tilde{f}(\lambda\,;\,\mu,\,\nu) = \int_{\mathbb{R}^2}f(q,p)\delta(\lambda - \mu q -\nu p)\,\mathrm{d}q \,\mathrm{d}p\,, 
\end{equation} 
which can be read as the expectation value of the family of observables $\delta(\lambda - \mu q -\nu p)$ with respect to the probability density $f$. There exists an inversion formula according to which it is possible to reconstruct the function $f$ knowing a sufficient number of tomograms, called a \textit{quorum}. In $\mathbb{R}^2$ the formula is a Fourier-like transform \cite{MMMS-2013} expressed as follows
\begin{equation}\label{Anti-radon transform}
f(q,p) = \frac{1}{(2\pi)^2}\int \tilde{f}(\lambda\,;\,\mu,\,\nu) \mathrm{e}^{i(\lambda - \mu q - \nu p)}\, \mathrm{d}\lambda\,\mathrm{d}\mu \,\mathrm{d}\nu\,.
\end{equation} 
Tomograms $\tilde{f}(\lambda\,;\,\mu,\,\nu)$ have the following properties:
\begin{itemize}
\item nonnegative $\tilde{f}(\lambda\,;\,\mu,\,\nu)\geq 0$;
\item integrable $\int \tilde{f}(\lambda\,;\,\mu,\,\nu) \,\mathrm{d}\lambda < \infty$ ;
\item homogeneous $\tilde{f}(c\lambda\,;\,\mu,\,\nu) = \frac{1}{c}\tilde{f}(\lambda;\,\mu,\,\nu)$, $\forall c > 0$.
\end{itemize}
The birth of Quantum Tomography, instead, is associated with a question raised by Pauli \cite{Pauli-1933}, and then it evolved towards the modern problem of Quantum State Tomography, as we have already mentioned in the introduction. From the point of view of exact tomography, the Weyl-Wigner formalism allows to obtain a reconstruction formula by introducing the operators
\begin{equation}
\delta(\lambda\mathbf{1} - \mu \hat{Q} - \nu \hat{P})\,, \quad \mathrm{e}^{i(\mu \hat{Q} + \nu \hat{P})}\,.
\end{equation} 
Then, given a state $\rho$, the corresponding tomograms are defined as 
\begin{equation}\label{eq.46}
\tilde{\rho}(\lambda\,;\,\mu\,,\,\nu) = \mathrm{Tr}(\rho\, \delta(\lambda\mathbf{1} - \mu \hat{Q} - \nu \hat{P})) 
\end{equation} 
and one can get the initial state $\rho$ via the following integral transform \cite{IMMS-2009}
\begin{equation}\label{eq.47}
\rho = \int \tilde{\rho}(\lambda\,;\,\mu,\,\nu) \mathrm{e}^{i(\lambda\mathbf{1} - \mu \hat{Q} - \nu \hat{P})}\, \mathrm{d}\lambda\,\mathrm{d}\mu \,\mathrm{d}\nu\,.
\end{equation}
Let us notice here that the tomograms in the quantum settings can be read also as the Radon transform of the Wigner function associated with a normal quantum state. However, while in the quantum case Wigner functions need not be nonnegative, the associated tomograms are still nonnegative functions and the inversion formula Eq.\eqref{Anti-radon transform} still holds. 

From a purely mathematical point of view Eqs.(\ref{eq.46},\ref{eq.47}) have to be understood only at a formal level and suitable domains for the operators and the distributions should be considered to get a rigorous formulation. To avoid these analyitical difficulties which are not essential for the development of the main idea of the paper, in the remaining sections we will focus on spin tomography where the quantum system possesses only a finite number of levels.  

\section{Spin tomography and the groupoid picture of Quantum Mechanics}\label{sec.4 spin tomography}
\subsection{Finite dimensional quantum tomography: the q-bit}
Spin tomography is concerned with finite dimensional quantum system. In this case, the Hilbert space which we consider is $\mathcal{H}=\mathbb{C}^n$, with the standard inner product, and the algebra of observables is given by the self-adjoint elements of the algebra $\mathcal{B}(\mathcal{H})=M_n(\mathbb{C})$, where $M_n(\mathbb{C})$ denotes the algebra of $n\times n$ matrices with complex entries. Given an observable $A \in \mathcal{B}(\mathcal{H})$, i.e., a Hermitean matrix, there is an orthonormal basis $\left\lbrace \left.|e_j\right\rangle \right\rbrace_{j=1,\cdots\,,n}$ of the Hilbert space $\mathcal{H}$ such that $A = \sum_{j=1}^n \lambda_j \left.|e_j\right\rangle \left\langle e_j | \right.$. Given this basis, it is possible to construct a matrix $E=(E_{jk})$ of rank-one operators $E_{jk}= \left.|e_j\right\rangle \left\langle e_k | \right.$, and for every normalized vector $\left.|\psi \right\rangle \in \mathcal{H}$, there is a corresponding pure state $\rho_{\psi} = \left.|\psi\right\rangle \left\langle \psi | \right.$. Using these two ingredients it is possible to define a new matrix with entries $\Psi_{jk} = \left\langle  \psi |e_j\right\rangle \left\langle e_k | \psi \right\rangle$, whose diagonal elements form a probability vector called the tomogram of $\rho_{\psi}$ with respect to $A$. The problem of exact spin tomography \cite{DM-1997,MM-1997,DMP-2003} amounts to reconstruct the state $\rho_{\psi}$ using a sufficient number of tomograms associated with a suitable family of orthonormal bases, or of observables $A$. As shown in \cite{MMSSS-2006}, in general such a procedure is successful if we use a family of projectors which form a total set in the Hilbert space $\mathcal{I}_2\subset \mathcal{B}(\mathcal{H})$ of Hilbert-Schmidt operators whose scalar product is $\left\langle  A | B \right\rangle = \mathrm{Tr}(A^{\dagger} B)$. In particular, when $\mathcal{H}$ is finite dimensional $\mathcal{I}_2 \cong \mathcal{B}(\mathcal{H})=M_n(\mathbb{C})$ a family of projectors $\left\lbrace P_{\mu} \right\rbrace_{\mu \in \mathcal{M}}$ forms a quorum if it determines a decomposition of the identity superoperator on $\mathcal{I}_2$.

Then, the reconstruction formula $\rho_{\psi} = \sum_{\mu\in M} \mathrm{Tr}(P_{\mu}\rho_{\psi}) P_{\mu}$ holds and the quantities $\mathrm{Tr}(P_{\mu}\rho_{\psi})$ are the components of the tomograms for this tomographic set of observables. It is obvious that the above reconstruction procedure still holds if one replaces pure states with generic mixed states.

As an example we can deal with the q-bit case, which corresponds to a two-dimensional quantum system. In such a case the Hilbert space is $\mathcal{H}=\mathbb{C}^2$ and we can consider a family of observables given by the Pauli matrices 
\begin{equation}
\sigma_1 = \left(  
\begin{array}{cc}
0 & 1 \\
1 & 0
\end{array}
\right)\,,\quad
\sigma_2 = \left(  
\begin{array}{cc}
0 & -i \\
i & 0
\end{array}
\right)\,,\quad
\sigma_3 = \left(  
\begin{array}{cc}
1 & 0 \\
0 & -1
\end{array}
\right)
\end{equation}
and the corresponding decomposing projectors contains the following tomographic set
\begin{equation}
\begin{split}
P_1 &= \left(  
\begin{array}{cc}
1 & 0 \\
0 & 0
\end{array}
\right)\,,\quad
P_2 = \left(  
\begin{array}{cc}
0 & 0 \\
0 & 1
\end{array}
\right)\, \\
P_3 &= \frac{1}{2}\left(  
\begin{array}{cc}
1 & 1 \\
1 & 1
\end{array}
\right)\,,\quad
P_4 = \frac{1}{2}\left(  
\begin{array}{cc}
1 & -i \\
i & 1
\end{array}
\right)\,.
\end{split}
\end{equation}
Indeed, this set forms a non orthonormal basis of the space $\mathcal{B}(\mathcal{H})=M_2(\mathbb{C})$ from which it is possible to get the othonormal basis given by the four operators
\begin{equation}
\begin{split}
W_1=P_1 &= \left(  
\begin{array}{cc}
1 & 0 \\
0 & 0
\end{array}
\right)\,,\quad
W_2 =P_2 =  \left(  
\begin{array}{cc}
0 & 0 \\
0 & 1
\end{array}
\right)\, \\
W_3 &= \left(  
\begin{array}{cc}
0 & 1 \\
0 & 0
\end{array}
\right)\,,\quad
W_4 = \left(  
\begin{array}{cc}
0 & 0 \\
1 & 0
\end{array}
\right)\,.
\end{split}
\end{equation}
Then, if we call $(p_1,1-p_1)$, $(p_2,1-p_2)$, $(p_3,1-p_3)$ the tomograms corresponding to the decompositions of $\sigma_1\,, \,\sigma_2\,,\,\sigma_3$, a state $\rho$ can be expressed as follows
\begin{equation}
\rho = \frac{1}{2}\mathbf{1} + \left( p_1-\frac{1}{2} \right)\sigma_1 + \left( p_2-\frac{1}{2} \right)\sigma_2 + \left( p_3-\frac{1}{2} \right)\sigma_3\,.
\end{equation} 
In the particular instance of a pure state, $\rho^2=\rho$, the following identity holds
\begin{equation}
\left( p_1-\frac{1}{2} \right)^2 + \left( p_2-\frac{1}{2} \right)^2 + \left( p_3-\frac{1}{2} \right)^2 = \frac{1}{4}
\end{equation}
whereas, mixed states lie in the interior of the (Bloch) sphere. On the other hand, uncorrelated probability vectors form a cube in the three-dimensional space $(p_1\,,\,p_2\,,\,p_3)$, so that quantum states are contained in a sphere within the cube of the uncorrelated triplets $(p_1\,,\,p_2\,,\,p_3)$.

From the perspective of Quantum State Tomography, on the other hand, one imposes additional requirements on the operators that are used to reconstruct any state. In particular, using unbiased measurements allows to maximize the information one gains about the ensamble performing n+1 measurements. In the case of q-bit tomography, a family of mutually unbiased measurements is made up of the measurements associated with the three Pauli matrices:
\begin{equation*}
\left\lbrace  
\left(
\begin{array}{c}
1 \\
0
\end{array}
\right)\,,\,
\left(
\begin{array}{c}
0 \\
1
\end{array}
\right)
\right\rbrace\,,\;
\left\lbrace  
\frac{1}{\sqrt{2}}\left(
\begin{array}{c}
1 \\
1
\end{array}
\right)\,,\,
\frac{1}{\sqrt{2}}\left(
\begin{array}{c}
1 \\
-1
\end{array}
\right)
\right\rbrace\,,\;  
\left\lbrace  
\frac{1}{\sqrt{2}}\left(
\begin{array}{c}
1 \\
i
\end{array}
\right)\,,\,
\frac{1}{\sqrt{2}}\left(
\begin{array}{c}
1 \\
-i
\end{array}
\right)
\right\rbrace\,.  
\end{equation*}   

On the other hand, it is possible to get reconstruction formulae also using SIC-POVMs, which are made up of $n^2$ positive operators multiple of rank-one projectors $P_j = \mid \psi_j \rangle  \langle  \psi_j  \mid$, where $n$ is the dimension of the finite dimensional Hilbert space under investigation. These operators form an informationally complete POVM when they are independent and form a basis of the space of all operators on the corresponding Hilbert space. The additional property of being symmetric means that $\mid \langle  \psi_j \mid \psi_k \rangle \mid^2 = 1/(n+1)$. Among the IC-POVM, the symmetric ones satisfy an additional condition: They form minimal spherical 2-designs (in the sense that have the minimum number of elements a 2-design can have) and minimize the second frame potential (see \cite{RBKSC-2004} for definition and proof of this result). For the q-bit case a SIC-POVM is obtained via the following operators \cite{REK-2004} depending on three-dimensional vectors $\overrightarrow{a}_j$, $j=1,\cdots,4$: 
\begin{equation}
P_j = \frac{1}{4}\left( \mathbf{1} + \overrightarrow{a}_j \cdot \overrightarrow{\sigma} \right)\,, \quad \overrightarrow{a}_j \cdot \overrightarrow{a}_k = \frac{4}{3}\delta_{jk} - \frac{1}{3}\,,\quad j,k=1,\cdots,4\,,
\end{equation}
where we denoted $\overrightarrow{\sigma} = (\sigma_1, \sigma_2, \sigma_3)$ the vector whose components are the three Pauli matrices.   

\subsection{Groupoid approach to Quantum Mechanics}
In the rest of this section we are going to revisit exact quantum tomography adopting a different formalism based on the use of groupoids and derived from Schwinger's picture of Quantum Mechanics \cite{CDIM-2020a}. Let us shortly recall the basic ingredients of this description: 
\begin{itemize}
\item A physical system is described in terms of a pair $\left( \mathcal{G}\,,\, [\mu] \right)$, where $\mathcal{G}\rightrightarrows \Omega$ is a groupoid and $[\mu]$ is an equivalence class of measures on the groupoid $\mathcal{G}$ (two measures are equivalent if they have the same sets of zero measure). 
\item The space of measurable functions is endowed with a convolution product $\star_{\mu}$ and an involution operation $\dagger$, making it a $\ast$-algebra. Choosing a suitable completion \cite{Hahn-1978} it is possible to make this algebra into a von-Neumann algebra $\mathcal{V}(\mathcal{G})$, called von-Neumann groupoid-algebra, whose real elements form the Jordan-Lie Banach algebra $\mathcal{O}$ of observables of the theory.
\item The states of the groupoid-algebra $\mathcal{V}(\mathcal{G})$, i.e., positive, linear and normalized functionals define the states of the physical system. 
\item Given two groupoids $\mathcal{G}_1 \rightrightarrows \Omega_1$ and $\mathcal{G}_2 \rightrightarrows \Omega_2$ there are two possible compositions in this setting: on one side there is the direct product of groupoids $\mathcal{G}_1\times \mathcal{G}_2 \rightrightarrows \Omega_1\times \Omega_2$, whereas on the other side we have the free product of groupoids $\mathcal{G}_1\ast \mathcal{G}_2 \rightrightarrows \Omega_{12}$, which depends, however, on a pair of additional inclusion maps $i_j\,\colon\,\Omega_j\,\rightarrow \,\Omega_{12}$, with $j=1,2$. These composition rules describe different physical notions of ``composition'' of systems, the first one being von Neumann's composition, while the second one reflects Lieb-Yngvason description of composition of thermodynamical systems \cite{LY-1999}. For more details we refer to \cite{CDIM-2020b}.
\item Concerning the evolution map, Schwinger's picture of Quantum Mechanics is characterized by an Action Principle which determines the propagator of the theory. It has been shown that such propagator can be determined from a q-Lagrangian, which is an observable in $\mathcal{V}(\mathcal{G})$ associated to a special family of states called DSF (Dirac-Schwinger-Feynman states)\cite{CDIMSZ-2021b}. On the other hand, one can define an evolution map also as a one-parameter group of automorphims of the groupoid-algebra $\mathcal{V}(\mathcal{G})$. 
\end{itemize}
For the sake of self-consistence, let us add some more details about finite connected groupoids which we will focus on for the rest of the paper. 

A finite groupoid $\mathcal{G}\rightrightarrows \Omega$ consists of a set $\Omega$ of objects, often representing observed outcomes of actual measurements on the system, and a set of morphisms $\alpha:x\rightarrow y$ with $x,y \in\Omega$, abstracting the idea of transitions of the system. Moreover, there is a pair of maps
\begin{equation}
s\,\colon\,\mathcal{G}\,\rightarrow \,\Omega\,,\quad t\,\colon\,\mathcal{G}\,\rightarrow\,\Omega\,,
\end{equation}   
associating each morphism $\alpha\,\colon\,x\rightarrow\,y$ with its source, $s(\alpha)=x$, and its target $t(\alpha)=y$. Transitions can be composed according to a partial composition law $\beta\circ\alpha$, defined whenever $t(\alpha)=s(\beta)$, which satisfies the following axioms:
\begin{itemize}
\item Associativity: $(\gamma\circ\beta)\circ\alpha = \gamma\circ(\beta\circ\alpha)$;
\item Units: For every outcome $x\in \Omega$ there is a transition, $1_x$, called the unit at $x$, which leaves invariant any other composable transition, i.e., $\alpha \circ 1_x = 1_y \circ \alpha = \alpha\,$;
\item Inverses: For every transition $\alpha\,\colon \,x\rightarrow \, y$, there exists an inverse transition $\alpha^{-1}\,\colon\,y\,\rightarrow\,x$, such that $\alpha\circ \alpha^{-1}=1_y$ and $\alpha^{-1}\circ\alpha = 1_x$ (this property implements Feynman's principle of ``microscopic reversibility''\cite{Feynman-2005}).
\end{itemize}
The set of transitions $\gamma\,\colon\,x\,\rightarrow\,x$ is the isotropy group of the object $x$ and it is denoted by $Aut(x)$. Since the groupoid is connected, i.e., for any $x,y\in \Omega$, there is $\alpha\,\colon\,x\,\rightarrow\,y$, the isotropy subgroups are all isomorphic to each other, thus we can identify all of them with an abstract group, say $\Gamma$. The isotropy subgroupoid, $\mathcal{G}_0 = \cup_{x\in\Omega} Aut(x)\subset \mathcal{G}$, is a normal subgroupoid such that the following exact sequence of morphisms of groupoids holds
\begin{equation}
\mathbf{1}_{\Omega}\rightarrow\mathcal{G}_0 \stackrel{i}{\to} \mathcal{G}\stackrel{\Pi}\rightarrow\mathcal{G}(\Omega)\rightarrow\mathbf{1}\,,
\end{equation}
where $\mathcal{G}(\Omega) = (t \times s)(\mathcal{G)}$ is a subgroupoid of the pair groupoid $\Omega \times \Omega$ of the set $\Omega$; since $\mathcal{G}$ is connected, $\mathcal{G}(\Omega) = \Omega \times \Omega$. The groupoid $\mathbf{1}_{\Omega}$ is made up only of the units and $\mathbf{1}$ is the trivial groupoid with only one object and one unit. The homomorphisms $i$ is the canonical embedding of $\mathcal{G}_0$ as a subgroupoid of $\mathcal{G}$ and the canonical projection $\Pi$ is the groupoid homomorphism $t \times s$. This sequence splits and there exists an isomorphism of groupoids such that $\mathcal{G}\cong \Gamma\times \mathcal{G}(\Omega)$ \cite{Ibort2019}. 

A finite dimensional quantum system (e.g., a quantum system with a finite number, say $n$, of energy levels) is described by the pair groupoid $\mathcal{G}(\Omega)\rightrightarrows \Omega$ over the set of integers $\Omega = \left\lbrace 1,\,2,\,\cdots,\,n \right\rbrace$. A transition is given by the pair $(k,j)\,\colon\,j\,\rightarrow\,k$ with $j,k=1,2,\cdots,n$. Two transitions $(m,l)$ and $(k,j)$ can be composed if $k=l$ and the resulting transition is $(m,k)\circ\,(k,j) = (m,j)$, which is the groupoidal expression of the Ritz-Rydberg combination principle of frequencies in spectral lines (something that was already noted by Connes in \cite{Connes-1994} chapter 1.1). Given this finite set we can construct an algebra following the procedure in Sec.\ref{Q-D_section}. Indeed, we have the set $\mathcal{V}(\mathcal{G})$ of all formal linear combinations of elements of the set $\mathcal{G}(\Omega)$, say $A_f=\sum_{j,k=1}^n f(k,j)(k,j)$, where $f$ is a complex-valued function on $\mathcal{G}(\Omega)$. This set forms a vector space of dimension $n^2$ isomorphic to $M_n(\mathbb{C})$, which is endowed with the following convolution product
\begin{equation*}
\begin{split}
A_f \star A_g &= \left( \sum_{m,l=1}^{n}f(m,l)(m,l) \right) \cdot  \left( \sum_{j,k=1}^{n}g(k,j)(k,j) \right) = \\
&=\sum_{j,k=1}^{n}\sum_{m,l=1}^{n}f(m,l)g(k,j)(m,l)\circ (k,j) = \\
= \sum_{j,k=1}^{n}\sum_{m,l=1}^{n}& f(m,l)g(k,j)\delta_{lk}(m,j) = \sum_{m,j=1}^{n}\left(\sum_{k=1}^{n}f(m,k)g(k,j)\right)(m,j) \,.
\end{split}
\end{equation*}
Realizing that $A_f$ can also be thought of as a square matrix, a direct check shows that $A_f \star A_g$ is nothing but the standard row-column product among matrices.  On the other hand, note that $A_f \star A_g = A_{f\ast g}$, where $(f \ast g)(m,j) = \sum_{k=1}^{n}f(m,k)g(k,j)$. An involution operation is given by the following rule $(A_f)^{\dagger} = \sum_{j,k=1}^n \overline{f(k,j)}(j,k)$, and the vector space $\mathcal{V}(\mathcal{G})$ endowed with the convolution product $\star$ and the involution $\dagger$ form an associative algebra which is isomorphic to $M_n(\mathbb{C})$. Any element of $\mathcal{V}(\mathcal{G})$ acts as an operator on the Hilbert space $\mathcal{H}=\mathcal{L}^2(\mathcal{G})$ via the left regular representation, i.e., $\pi\,\colon\,\mathcal{V}(\mathcal{G})\,\rightarrow\,\mathcal{B}(\mathcal{H})$ where $\pi(A_f) \psi = f \ast \psi$. The weak-closure (which for a finite groupoid is equivalent to the norm-closure) of $\pi(\mathcal{V}(\mathcal{G}))$ is the groupoid-algebra of the system and it is isomorphic to the Von-Neumann algebra $M_n(\mathbb{C})$, which is a finite type-I factor. A state of the system is a normalized positive linear functional, say $\omega$, on $\mathcal{V}(\mathcal{G})$ and, in this discrete case, it can be identified with the function $\varphi(k,j) = \omega(\delta_{(k,j)})$, where $\delta_{(k,j)}$ is the function on $\mathcal{G}(\Omega)$ which is zero except at the transition $(k,j)$. For such a factor there exists a tracial state \cite{Blackadar-2006} $\tau$ whose action on an element of the algebra $\mathcal{V}(\mathcal{G})$ reads as
\begin{equation}
\tau(A_f) = \frac{1}{n}\sum_k f(k,k)\,. 
\end{equation}
Using such tracial state any other state can be written as
\begin{equation}\label{normal_states}
\omega (A_f) = \tau (A_{\varphi}^{\dagger} \star A_f)= \tau (A_{\varphi} \star A_f)  = \frac{1}{n}\sum_{(j,k)\in \mathcal{G}} \overline{\varphi(j,k)} f(j,k)\,,
\end{equation}
where $\varphi$ is a positive semi-definite function on the groupoid, i.e., for any $N\in \mathbb{N}$, $\xi_r \in \mathbb{C}$ and $\alpha_r \in \mathcal{G}$, $r,s=1,\cdots\,,N$, with $t(\alpha_s)=t(\alpha_r)$, one has that: 
\begin{equation}\label{semi-definite positive functions}
\sum_{r,s=1}^N \overline{\xi}_r \xi_s \varphi(\alpha_r^{-1}\circ \alpha_s) \geq 0
\end{equation} 

Under this identification of states and positive semi-definite functions on $\mathcal{G}$, the evaluation of the state $\omega$ on the element of the groupoid-algebra $A_f$ reads as
\begin{equation}\label{normal state}
\omega(A_f) = \tau (A_{\varphi}^{\dagger} \star A_f) = \frac{1}{n}\left\langle \varphi , f \right\rangle_{\mathcal{L}^2(\mathcal{G})}\,,
\end{equation}
that is, it is the normalized scalar product on $\mathcal{L}^2(\mathcal{G})$ of $\varphi$ and $f$. 

Before coping with the problem of tomographic reconstruction, a last ingredient is needed: A \textit{bisection} is a subset $b\subset \mathcal{G}$ such that the restriction of the source and target maps to it are bijections (see \cite{CDIMS-2021} for more details in the finite case and \cite{Mackenzie-2005} for the smooth situation). Therefore, to any bisection $b$ there are associated two sections 
\begin{equation}
b_s\,:\,\Omega\,\rightarrow\,\mathcal{G}\,,\quad b_t\,\colon\,\Omega\,\rightarrow\,\mathcal{G}\,,
\end{equation} with respect to the source and target maps, respectively, i.e., $s\circ b_s = \mathrm{id}_\Omega$, $t\circ b_t = \mathrm{id}_\Omega$. The set of all bisections, we will call it $\mathcal{S}_n$, form a group under the (canonical) multiplication law:
\begin{equation}
(b'\cdot b)_s (x) = b'_s(t(b_s(x))) \circ b_s(x)\,, 
\end{equation}
where $x\in \Omega$, and an analogous formula can be written for the section $(b'\cdot b)_t$. In the case of the pair-groupoid $\mathcal{G}(\Omega)$ over a finite set the group of bisections is isomorphic to the group $\mathbb{S}_n$ of all permutations of the elements in $\Omega$. This group will play a fundamental role in the rest of the section.  

\subsection{Spin Tomography and frames on $\mathcal{G}(\Omega)$}\label{sec 4.3}

Concerning the tomographic reconstruction procedure, it can be divided into two steps: the first one consists in associating a family of tomograms to any state $\omega$ and the second one consists in applying a suitable reconstruction formula using the tomograms. The notion of frames helps in both steps. In what follows we will shortly recall the main features of the theory of frames, mainly referring to \cite{IL-2021} (for more details see \cite{Daubechies-1992,Kaiser-1994}). 

Given a complex separable Hilbert space $\mathcal{H}$ and a measurable space $\mathcal{M}$ endowed with a measure $\mu$, a family of vectors $\mathcal{F}=\left\lbrace \left. \mid \psi_x \right\rangle \,, \,\mid x\in\mathcal{M} \right\rbrace$ is called a frame based on $\mathcal{M}$ if it satisfies:
\begin{itemize}
\item For every $\left.|\psi\right\rangle $ the function $\mathrm{ev}_{\psi}(x) = \left\langle \psi \mid \psi_x \right\rangle$ is $\mu$-measurable and belongs to $\mathcal{L}^2(\mathcal{M},\mu)$;
\item There are real numbers $0 < A \leq B$ such that
\begin{equation}
A\parallel \psi \parallel^2 \leq \parallel \mathrm{ev}_{\psi} \parallel^2_{\mathcal{L}^2(\mathcal{M})} = \int_{\mathcal{M}} \mid \left\langle \psi , \psi_x \right\rangle \mid^2 \mathrm{d}\mu (x) \,\leq B \parallel \psi \parallel^2 \,,
\end{equation}
for all $\left. \mid \psi \right\rangle \in \mathcal{H}$. In particular, when the constants $A$ and $B$ can be chosen to be equal, the frame is called tight. 
\end{itemize}
Given a frame $\mathcal{F}$ one defines a bounded linear operator, called the frame operator
\begin{equation}
\mathfrak{F}\,\colon\, \mathcal{H} \, \rightarrow \, \mathcal{L}^2(\mathcal{M})\,, \quad  \mathfrak{F}|\psi \rangle = \langle \psi_{( \cdot )} \mid \psi \rangle\,.  
\end{equation}
The frame operator is injective and it admits a bounded left inverse. The adjoint frame operator is then defined as follows
\begin{equation}
\begin{split}
\mathfrak{F}^*\,\colon\, \mathcal{L}^2(\mathcal{M})\, \rightarrow\, \mathcal{H}\,, \quad  &\langle \mathfrak{F}^*f |\psi \rangle_{\mathcal{H}} = \langle f , \mathfrak{F}|\psi\rangle \rangle_{\mathcal{L}^2(\mathcal{M})}\\
\mathfrak{F}^*f =& \int_{\mathcal{M}} f(x)|\psi_x\rangle \mathrm{d}\mu (x)\,.
\end{split}
\end{equation}
Using the frame operator and its adjoint one can define the metric operator $S = \mathfrak{F}^*\mathfrak{F}\,\colon \, \mathcal{H}\,\rightarrow\,\mathcal{H}$ which is bounded and definite positive, so that it admits an inverse $S^{-1}$ which is still positive definite and bounded. Using the metric operator one can introduce another frame, called the dual frame, as follows
\begin{equation}
\mathcal{F}^* = \left\lbrace |\psi^x \rangle = S^{-1}| \psi_x \rangle \mid x\in \mathcal{M} \right\rbrace\,,
\end{equation}
and the following property holds, which will play a major role in the rest of this section
\begin{equation}
\mathbb{I} = \int_{\mathcal{M}} |\psi^x\rangle \langle \psi_x | \, \mathrm{d}\mu (x) = \int_{\mathcal{M}} |\psi_x\rangle \langle \psi^x | \, \mathrm{d}\mu (x)\,.
\end{equation}
Therefore, a reconstruction formula for any vector $|\psi \rangle \in \mathcal{H}$ can be written as follows
\begin{equation}
|\psi \rangle = \int_{\mathcal{M}} \left(\mathfrak{F}|\psi\rangle \right)(x)\,|\psi^x\rangle \, \mathrm{d}\mu(x) \,.
\end{equation}
Then, summarizing the above digression, a tomographic reconstruction formula can be obtained once a frame is introduced in the Hilbert space $\mathcal{L}^2(\mathcal{G})$, the tomograms being derived from the function $\mathfrak{F}|\psi\rangle $. In the rest of this section we will show that actually the group of the bisections will provide us with a frame. Moreover, when the finite set $\Omega$ is a finite field, a subgroup of the whole group of bisection, namely the group of affine liner transformation on $\Omega$, will determine a frame, too. Let us mention, however, that due to the peculiarity of the permutation group of two elements the construction which we will present cannot be applied to the q-bit case (i.e., to the case in which $\Omega$ is a 2-element set), where the procedures outlined at the beginning of this section are optimal, as we have already discussed. 

Given a bisection $b \subset \mathcal{G}(\Omega)$, the corresponding characteristic function $\chi_b$ defines a square-integrable function with respect to the counting measure on the groupoid and the associated element in $\mathcal{V}(\mathcal{G})$ is
\begin{equation}
A_b = \sum_{(k,j)\in \mathcal{G}(\Omega)}\chi_b(k,j)\,(k,j) = \sum_{j\in \Omega} (t(b_s(j)), j) = \sum_{j \in \Omega} (\sigma_b(j), j)\,,
\end{equation}
where $\sigma_b$ is a bijective map on $\Omega$, i.e., a permutation. Then, the set of all $A_b$, where $b$ is a bisection, defines a representation of the permutation group $\mathbb{S}_n$ on the Hilbert space $\mathcal{L}^2(\Omega)\cong \mathbb{C}^n$ which is isomorphic to the fundamental one, and so it is unitary. A feature of this representation which we will use next, is the fact that it is not irreducible, but for $n>2$ it is the direct sum of the trivial representation and the $(n-1)$-dimensional irreducible representation \cite{Cameron1999}. Moreover, let us notice that the scalar product on $\mathcal{L}^2(\mathcal{G})$ can be expressed as
\begin{equation}
\langle \psi_1 | \psi_2 \rangle_{\mathcal{L}^2(\mathcal{G})} = \sum_{j \in \Omega}\sum_{k \in \Omega} \overline{\psi_1(j,k)}\psi_2(j,k) = \mathrm{Tr}_{n}(\Psi_1^{\dagger}\Psi_2)\,, 
\end{equation}
where $\mathrm{Tr}_n$ is the standard trace on $n\times n$ matrices, and $(\Psi_a)_{j,k} = \psi_a(j,k)$, $a=1,2$, is a $n\times n$ matrix with complex entries (we have used the isomorphism between $\mathcal{V}(\mathcal{G})$ and $M_n(\mathbb{C})$).
In order to write down a tomographic reconstruction formula, we need the following result. 

\begin{theorem}
Let $\mathcal{G}(\Omega)$ be the groupoid of pairs over $\Omega$, a finite set of cardinality $n>2$. The set 
\begin{equation}
\mathcal{F}=\left\lbrace \chi_b \mid b \in \mathcal{S}_n \right\rbrace
\end{equation}
is a frame for the Hilbert space $\mathcal{H}=\mathcal{L}^2(\mathcal{G})$ over the group of bisections $\mathcal{S}_n$ equipped with the counting normalized measure $\sharp$. 
\end{theorem} 
\begin{proof*}
Since the group $\mathcal{S}_n$ is finite and the Hilbert space $\mathcal{H}$ is finite dimensional, for every $\mid \psi \rangle \in \mathcal{H}$ the map $\mathrm{ev}_{\psi}$ is $\mu$-measurable and belongs to $\mathcal{L}^2(\mathcal{S},\sharp)$. 

Then we have to prove that there are two real numbers $0< A\leq B$ such that 
\begin{equation}
A\parallel \psi \parallel^2 \leq \sum_{b\in\mathcal{S}_n}\, \frac{1}{n!} \; |\: \langle \psi | \chi_b \rangle_{\mathcal{L}^2(\mathcal{G})}\:|^2  \,\leq B \parallel \psi \parallel^2\,.
\end{equation}
The second inequality involving the constant $B$ is a consequence of the Cauchy-Schwarz inequality. Using the previous scalar product it is easy to see that $B=n^2$. In order to prove the first inequality let us notice that
\begin{equation}
\begin{split}
\sum_{b\in\mathcal{S}_n}\, \frac{1}{n!} \; |\: \langle \psi | \chi_b &\rangle_{\mathcal{L}^2(\mathcal{G})}|^2 = \sum_{b\in\mathcal{S}_n}\, \frac{1}{n!} \;  \langle \psi | \chi_b \rangle_{\mathcal{L}^2(\mathcal{G})} \langle \chi_b | \psi \rangle_{\mathcal{L}^2(\mathcal{G})} = \\
=& \sum_{b\in\mathcal{S}_n}\frac{1}{n!}\, \mathrm{Tr}_{n}(\Psi^{\dagger} U(b)) \mathrm{Tr}_{n}(U(b)^{\dagger}\Psi)\,,
\end{split}
\end{equation}
where $U(b)$ denotes the matrix associated with the bisection $b$ in the fundamental representation of the group $\mathbb{S}_n$. Let us now introduce the projector $P$ onto the one-dimensional vector subspace of $\mathbb{C}^n$ supporting the trivial representation of $\mathbb{S}_n$. Then we have
\begin{equation}
\begin{split}
&\frac{1}{n!} \sum_{b\in\mathcal{S}_n}\, \mathrm{Tr}_{n}(\Psi^{\dagger} U(b)) \mathrm{Tr}_{n}(U(b)^{\dagger}\Psi) = \\
&= \frac{1}{n!} \sum_{b\in\mathcal{S}_n}\, \left[ \mathrm{Tr}_{n}(\Psi^{\dagger} P U(b) P) + \mathrm{Tr}_{n}(\Psi^{\dagger} (\mathbf{I}-P) U(b) (\mathbf{I} - P))\right]\cdot \\
&\cdot\left[ \mathrm{Tr}_{n}(P U(b)^{\dagger} P \Psi) + \mathrm{Tr}_{n}((\mathbf{I}-P) U(b)^{\dagger} (\mathbf{I} - P) \Psi)\right] =\\
&= \frac{1}{n!} \sum_{b\in\mathcal{S}_n}\, \mathrm{Tr}_{n}(\Psi^{\dagger} P U(b) P)\mathrm{Tr}_{n}(P U(b)^{\dagger} P \Psi) + \\
&\frac{1}{n!} \sum_{b\in\mathcal{S}_n}\, \mathrm{Tr}_{n}(\Psi^{\dagger} (\mathbf{I} - P) U(b) (\mathbf{I} - P))\mathrm{Tr}_{n}((\mathbf{I} - P) U(b)^{\dagger} (\mathbf{I} - P) \Psi) = \\
&= \mathrm{Tr}_n(P\Psi^{\dagger}P\Psi) + \frac{1}{n-1}\mathrm{Tr}_n((\mathbf{I} - P) \Psi^{\dagger} (\mathbf{I} - P) \Psi) > \frac{1}{n-1}\parallel \psi \parallel^2_{\mathcal{L}^2(\mathcal{G})}\,,
\end{split}
\end{equation}
where in the second equation we have used the orthogonality properties of the irreducible unitaries representations of the permutation group for $n>2$ \cite{Ibort2019}. Since $A=\frac{1}{n-1}>0$ the family $\mathcal{F}$ is a frame, but it is not tight. 
\end{proof*}

In order to find the dual frame we need the inverse of the metric operator $S$. It can be easily computed using the same chain of equalities in the proof of Th.1:
\begin{equation}
\begin{split}
\langle \psi_1 |S|\psi_2&\rangle_{\mathcal{L}^2(\mathcal{G})} = \frac{1}{n!} \sum_{b\in\mathcal{S}_n}\, \mathrm{Tr}_{n}(\Psi_1^{\dagger} U(b)) \mathrm{Tr}_{n}(U(b)^{\dagger}\Psi_2) =\\
\frac{1}{n-1}&\left[ \mathrm{Tr}_n(\Psi_1^{\dagger}\Psi_2) + (d-2) \mathrm{Tr}_n(P\Psi_1^{\dagger}P\Psi_2) \right]\,,
\end{split}
\end{equation} 
so that 
\begin{equation}
S = \frac{1}{d-1}\left[ \mathbb{I} + (d-2)\Pi \right]
\end{equation}
where $\Pi\,\colon\,\mathcal{L}^2(\mathcal{G})\,\rightarrow \, \mathcal{L}^2(\mathcal{G})$ is the projection operator
\begin{equation}
(\Pi \psi)(k,j)  = \sum_{l\in\Omega}\sum_{m\in\Omega} \, P(k,l)\psi(l,m)P(m,j) \,.
\end{equation}
Since $\Pi$ is a projection operator, the inverse of the metric operator exists and it is equal to 
\begin{equation}
S^{-1} = (d-1)\left[ \mathbb{I} - \frac{(d-2)}{(d-1)} \Pi \right]\,.
\end{equation}
Then, the dual frame is written as
\begin{equation}
\mathcal{F}^* = \left\lbrace |\chi^b\rangle = S^{-1} |\chi_b \rangle \;| \; b \in \mathcal{S}_n \right\rbrace\,,
\end{equation}
and we have that 
\begin{equation}
\mathbb{I} = \frac{1}{n!}\sum_{b\in\mathcal{S}_n} |\chi_b\rangle \langle \chi^b |\,.
\end{equation}
Therefore, we can write a state as follows
\begin{equation}\label{reconstruction_formula}
\begin{split}
\omega(A_f) = \frac{1}{n}\langle \varphi | f \rangle_{\mathcal{L}^2(\mathcal{G})} &= \frac{1}{n!}\sum_{b\in \mathcal{S}_n}\frac{1}{n} \langle \varphi | \chi_b \rangle_{\mathcal{L}^2(\mathcal{G})}\langle \chi^b | f \rangle_{\mathcal{L}^2(\mathcal{G})}=\\
=& \frac{1}{n!}\sum_{b\in \mathcal{S}_n} F_{\varphi}(b) \langle \chi^b | f \rangle_{\mathcal{L}^2(\mathcal{G})} \,, 
\end{split}
\end{equation}
and if one looks more carefully at the term $F_{\varphi}(b)$ one can notice that it can be rewritten as follows
\begin{equation}
F_{\varphi}(b) = \frac{1}{n} \langle \varphi | \chi_b \rangle_{\mathcal{L}^2(\mathcal{G})} = \frac{1}{n}\mathrm{Tr}_n (\Phi U(b)) = \sum_{m= 1}^n \frac{e^{i\theta_m(b)}}{n} \langle m |\Phi | m\rangle_{\mathbb{C}^n}\,,
\end{equation}
where $ | m \rangle $ is a basis of eigenfunctions of the matrix $U(b)$ with eigenvalues $\mathrm{e}^{i\theta_m(b)}$. The vector with components
\begin{equation}
p_m = \frac{1}{n} \langle m |\Phi | m\rangle_{\mathbb{C}^n}
\end{equation}
is a probability vector, which is the tomogram associated via the bisection $b$ to the state $\Phi$. As a final comment, let us notice that the function $F_{\varphi}(b)$ can be expressed in terms of the scalar product 
\begin{equation}
F_{\varphi}(b) = \frac{1}{n} \langle \varphi | \chi_b \rangle_{\mathcal{L}^2(\mathcal{G})} = \frac{1}{n} \sum_{(k,j)\in \mathcal{G}} \overline{\varphi}(j,k)\chi_b(j,k)\,,
\end{equation}
and since $\chi_b$ is a characteristic function, the previous formula can be interpreted as the ``integral'' over the subset $b$ which provides a clear analogy with the classical Radon transform. 

Such analogy becomes even more evident if one consider the case in which $\Omega$ is the set $\mathbb{Z}_n$, of integers modulo $n$, with $n$ an odd prime number, such that $\Omega$ is a finite field. In this case, one can consider a subgroup of the group of all permutation, which is the group $\mathrm{Aff}_n$ of affine linear transformation, an element of which is denoted by the pair $(\mu,l) = g$, with $\mu, l \in \Omega$, $\mu \neq 0$. This group acts on the set $\Omega$ itself as $\sigma_{(\mu, l)} (j) = \mu j + l$, and the action is transitive, since given $k,j\in \Omega$ there exists $(\mu,l)\in \mathrm{Aff}_n$ such that $k = \mu j + l$ (take, for instance, $\mu=1$ and $l=k-j$). With an abuse of notation the set of bisections associated with the graph of this action will be denoted by $\mathrm{Aff}_n$. Its elements, analogously, will be denoted by the symbol $g\subset \mathcal{G}(\Omega)$ and the corresponding characteristic functions by $\chi_g$. The matrices associated with these characteristic functions will be written as $U(g)$ and form a representation of the group $\mathrm{Aff}_n$ which is unitary and not irreducible. However it is the direct sum of the trivial representation and the unique $n-1$-dimensional irreducible representations \cite{Huppert1998}. This can be proven by looking at the characters of the irreducible representation of the group $\mathrm{Aff}_n$ and noticing that 
\begin{equation}
\begin{split}
\mathrm{Tr}(U(g)) = 0\quad \quad \mathrm{for}\,\,g = (1,l)\\
\mathrm{Tr}(U(g)) = 1\quad \quad \mathrm{for}\,\,g = (\mu,l),\;\mu\neq 1
\end{split}  
\end{equation}
where the last equality comes from the fact that the equation $\sigma_{(\mu,l)}(j) = j$ has a unique solution, $j=-\frac{l}{\mu -1}$ and consequently there is a unique fixed point under the action defined by $\sigma$.

Now the same proof as in Th.1 can be applied to show that the sets
\begin{equation}
\begin{split}
&\mathcal{F}=\left\lbrace |\chi_g\rangle \mid g \in \mathrm{Aff}_n \right\rbrace \\
\mathcal{F}^* =& \left\lbrace |\chi^g\rangle = S^{-1} |\chi_g \rangle \;| \; g \in \mathrm{Aff}_n \right\rbrace\,,
\end{split}
\end{equation}
form a pair of dual frames which can be used to provide the following tomographic reconstruction formula 
\begin{equation}
\begin{split}
\omega(A_f) = \frac{1}{n}\langle \varphi | f \rangle_{\mathcal{L}^2(\mathcal{G})} &= \frac{1}{n^2(n-1)}\sum_{g\in \mathrm{Aff}_n} \langle \varphi | \chi_g \rangle_{\mathcal{L}^2(\mathcal{G})}\langle \chi^g | f \rangle_{\mathcal{L}^2(\mathcal{G})} =\\
=&\frac{1}{n(n-1)}\sum_{g\in \mathrm{Aff}_n} F_{\varphi}(g) \langle \chi^g | f \rangle_{\mathcal{L}^2(\mathcal{G})} \,. 
\end{split}
\end{equation}
In this case the correspondence with the classical Radon transform for $\mathbb{R}\times\mathbb{R}$ is more evident, since the subset $g$ is the analogue of a linear submanifold of the plane in a discrete and finite setting and, accordingly, the term $F_{\varphi}(g)$ can be interpreted as the ``integral'' over this linear submanifold. Moreover, we have that, for $g=(1,l)$, it holds
\begin{equation*}
\label{fourier transform}
\begin{split}
&n F_{\varphi}((\mu=1, l)) = \sum_{(k,j)\in \mathcal{G}} \overline{\varphi}((k,j))\delta(k-j-l) = \\ 
&=\frac{1}{n}\sum_{(k,j)\in \mathcal{G}} \sum_{m=0}^{n-1} \overline{\varphi}((k,j)) \exp\left( i\frac{2\pi}{n}m(k-j-l) \right) = \\
&=\sum_{m=0}^{n-1} e^{-i\frac{2\pi}{n}ml}\sum_{(k,j)\in \mathcal{G}} \overline{\varphi}((k,j))\frac{e^{-i\frac{2\pi}{n}mj}}{\sqrt{n}}\frac{e^{i\frac{2\pi}{n}mk}}{\sqrt{n}} =\\ &=\sum_{m=0}^{n-1} e^{-i\frac{2\pi}{n}ml}\sum_{(k,j)\in \mathcal{G}} \overline{\varphi}((k,j)) \psi_m(k) \overline{\psi}_m(j) = \\
&=\sum_{m=0}^{n-1} e^{-i\frac{2\pi}{n}ml} \langle m |\Phi | m\rangle_{\mathbb{C}^n}
\end{split}
\end{equation*}
where $\delta$ denotes the function which is 1 whenever its argument is 0. From the above equation we have that the function $F_{\varphi}(1,l)$ is the discrete Fourier transform of the tomogram associated with $\varphi$ via the bisection $g$, and this is in agreement with Bochner's theorem \cite{Folland-2016}. On the other hand, if $g=(\mu,l)$ with $\mu\neq 1$ we can replace the indices $k,j$ in Eq.\eqref{fourier transform} with the new indices $r,s$ both ranging in $\Omega$ such that 
\begin{equation}
j= -\frac{l}{\mu-1} +r\,, \quad\quad k = -\frac{l}{\mu-1} +s\,.
\end{equation}
In this way we can write 
\begin{equation*}
\begin{split}
n F_{\varphi}((\mu, l)) = \sum_{(k,j)\in \mathcal{G}} \overline{\varphi}((\frac{-l}{\mu-1} +& s,\frac{-l}{\mu-1} +r))\chi_g(\frac{-l}{\mu-1} +s,\frac{-l}{\mu-1} +r)) = \\ 
=\varphi((\frac{-l}{\mu-1},\frac{-l}{\mu-1})) + \sum_{(r,s)\neq 0}& \sum_{m=1}^{n-1} \overline{\varphi}((\frac{-l}{\mu-1} +s,\frac{-l}{\mu-1} +r)) \delta(s-\mu r) = \\
=\varphi((\frac{-l}{\mu-1},\frac{-l}{\mu-1})) + \sum_{(r,s)\neq 0}& \sum_{m=1}^{n-1} \overline{\varphi}((\frac{-l}{\mu-1} +s,\frac{-l}{\mu-1} +r))\cdot \\
&\cdot \delta(\log(s)-\log(r)-\log(\mu)) =\\
=\varphi((\frac{-l}{\mu-1},\frac{-l}{\mu-1})) + \sum_{(r,s)\neq 0}& \sum_{m=1}^{n-1} \overline{\varphi}((\frac{-l}{\mu-1} +s,\frac{-l}{\mu-1} +r)) \cdot \\
&\cdot \frac{\exp\left( i\frac{2\pi}{n}m(\log(s)-\log(r)-\log(\mu)) \right)}{n-1}=\\
=\varphi((\frac{-l}{\mu-1},\frac{-l}{\mu-1})) +&\sum_{m=1}^{n-1}e^{-i\frac{2\pi}{n-1}m\log(\mu)} \sum_{(r,s)\neq 0}  \overline{\varphi}((\frac{-l}{\mu-1} +s,\frac{-l}{\mu-1} +r)) \cdot \\
&\cdot\frac{e^{-i\frac{2\pi}{n-1}m\log(r)}}{\sqrt{n-1}}\frac{e^{i\frac{2\pi}{n-1}m\log(s)}}{\sqrt{n-1}} = \\
=\varphi((\frac{-l}{\mu-1},\frac{-l}{\mu-1})) &+\sum_{m=1}^{n-1}e^{-i\frac{2\pi}{n-1}m\log(\mu)} \langle m_{l/\mu} |\Phi | m_{l/\mu} \rangle_{\mathbb{C}^n} = \\ =\sum_{m=0}^{n-1}e^{-i\frac{2\pi}{n-1}m\log(\mu)}& \langle m_{l/\mu} |\Phi | m_{l/\mu} \rangle_{\mathbb{C}^n}\,,
\end{split}
\end{equation*}
where $\log$ denotes the discrete logarithm which exists whenever $\Omega$ is a finite field. Therefore, also in this case the function $F_{\varphi}(\mu,l)$ is the discrete Fourier transform of the tomogram associated with $\varphi$ via the bisection $g$. Let us notice that using the group of affine linear transformations we found $n+1$ different orthonormal basis of the vector space $\mathbb{C}^n$ 
\begin{equation}
\left\lbrace |m\rangle \right\rbrace\,,\;\left\lbrace |m_{l/\mu}\rangle \;\mid\; l=0,1,\cdots,n-1 \rangle \right\rbrace
\end{equation}
but they are not mutually unbiased. 

The concept of mutually unbiased measurements, indeed, is connected to the concept of complementarity and in this sense the Heisenberg-Weyl group seems to plays a more important role, rather than the affine one. A family of mutually unbiased bases can be obtained as follows. Let us consider two operators $X$, $Y$ satisfying $XY = \omega YX$, where $\omega$ is a $n$-root of the identity. Then the eigenbases of the operators $X$, $Y$, $XY^{k}$, $k=1,\cdots , n-1$, are mutually unbiased. If we look at the characteristic functions supported on the graph of bisections, we can obtain the same eigenbases if we consider the DSF functions
\begin{equation}
V_r(k,j) = \exp(i(S_r(k)-S_r(j)))\,,\quad S_r(j) = \frac{r}{2}j(j-1)\,,\;r=1,\cdots,n-1
\end{equation}   
and the operators associated with the functions on the groupoid 
\begin{equation}
\tilde{U}(r)(k,j) = (V_r \chi_{(1,1)})(k,j) = \exp(i(S_r(j)-S_r(k))) \delta (k-j-1)\,.
\end{equation}
The new functions $\tilde{U}(r)$ are unitarily equivalent representations of the unit translation operator, and their powers provide representations of the group of integers modulo $n$, under the addition operation. Due to the connection between DSF functions and the variational principle in Quantum Mechanics, further work is required to understand whether some features of mutually unbiased bases could be derived directly from some variational problem. We will address this issue in forthcoming works. 

On the other hand, $POVM$s are a generalization of the concept of quantum measurements not involving orthogonal projectors. In the groupoid approach to Quantum Mechanics we have presented in this work, pieces of information about the state have been obtained by looking at bisections, leading us to families of orthonormal bases. According to the authors' point of view, a generalized procedure which would allow to include $POVM$s in the picture should be based on a notion of inference procedure which roots back at Cencov's theory \cite{Cencov-2000} on statistical decision rules. Enter the details of this formulation is far beyond the scope of this work and it will be addressed somewhere else. 

As a final remark let us notice that given a family of probability distributions we are not ensured that the function we are going to build using the reconstruction formula Eq.\eqref{reconstruction_formula} will define a quantum state. However, an answer to this question can be obtained by looking back at Eq.\eqref{normal_states}-\eqref{semi-definite positive functions}, where we observed that a quantum state is associated with a semi-definite positive function on the groupoid $\mathcal{G}(\Omega)$. Therefore, a family of probability distributions $\mathcal{P}= \left\lbrace p^{(b)}_m\,,\;m=0,1,\cdots,n-1 | \: b\in \mathcal{S} \right\rbrace$ can be interpreted as a family of tomograms of a quantum state iff the function 
\begin{equation}
\varphi = \frac{1}{n!}\sum_{b\in\mathcal{S}} \left( \sum_{m=0}^{n-1} p^{(b)}_m e^{i\theta_m(b)}\right) |\chi^b\rangle \,,
\end{equation}
is a semi-definite positive function on the groupoid $\mathcal{G}(\Omega)$. 

\section{Conclusions}
In this paper we have started to address the problem of Quantum State Tomography from the point of view of Schwingers picture of Quantum Mechanics based on groupoids and their algebras. In particular, this should be considered as a preliminary work where some questions related to exact tomography for finite dimensional quantum systems have been investigated. We have presented a tomographic reconstruction procedure for states of the groupoid-algebra of the pair groupoid $\mathcal{G}(\Omega)$ over a finite set $\Omega$. This procedure is inspired by the classical Radon transform, since tomograms are associated to the ``integral'' over discrete analogues of the graphs of homeomorphisms. In the particular instance of $n$ being an odd prime number, the set $\Omega$ of integers modulo $n$ is a field and we can restrict the integration to the discrete analogue of Lagrangian subspaces, which are images of the field $\Omega$ via affine linear maps. This point of view can be immediately generalized to the case where $n$ is a prime power, such that one can consider $\Omega$ to be the corresponding finite Galois field. Then, we can consider again the group of affine linear transformations. However, another question arises: in the case where $\Omega$ is not a field, is it possible to select a subset of bisections which would provide us with a frame and a reconstruction procedure? This would correspond to a reconstruction procedure where, instead of considering the integration over the graph of linear maps, one is taking the graphs of non-linear functions (the continuous analogue has been analyzed in \cite{AFMMPS-2012} for instance). On the other hand, Real State Tomography is a currently very active research field due to its fundamental role in the development of reliable Quantum Technologies. In conclusion, according to the autors' point of view, a deeper investigation of the problem of real tomography in the groupoidal framework of Schwinger's picture of Quantum Mechanics requires a formulation of inference and statistical decision in said context, and is postponed to future works.     

\section*{Acknowledgments}

The authors acknowledge financial support from the Spanish Ministry of Economy and Competitiveness, through the Severo Ochoa Programme for Centres of Excellence in RD (SEV-2015/0554), the MINECO research project  PID2020-117477GB-I00,  and Comunidad de Madrid project QUITEMAD++, S2018/TCS-A4342.
GM would like to thank partial financial support provided by the Santander/UC3M Excellence  Chair Program 2019/2020, and he is also a member of the Gruppo Nazionale di Fisica Matematica (INDAM), Italy. 
FDC acknowledges support from the CONEX-Plus programme funded by Universidad Carlos III de Madrid and the European Union's Horizon 2020 research and innovation programme under the Marie Sklodowska-Curie grant agreement No. 801538. 
FMC acknowledges that the work has been supported by the Madrid Government (Comunidad de Madrid-Spain) under the Multiannual Agreement with UC3M in the line of ``Research Funds for Beatriz Galindo Fellowships'' (C$\setminus$\&QIG-BG-CM-UC3M), and in the context of the V PRICIT (Regional Programme of Research and Technological Innovation).  

The authors would like to thank the anonymous referees for many comments which lead to a substantial improvement of the manuscript.

\end{document}